\documentstyle[12pt]{article}

\newcommand{\be}{\begin{equation}}
\newcommand{\ee}{\end{equation}}
\newcommand{\bea}{\begin{eqnarray}}
\newcommand{\eea}{\end{eqnarray}}
\newcommand{\ci}{\cite}
\newcommand{\bi}{\bibitem}
\newcommand{\nono}{\nonumber \\}
\newcommand{\tr}{{\rm tr}}

\newcommand{\e}{{\rm e}}

\newcommand{\ssF}{{\sin^2 F}}

\newcommand{\da}{\dagger}
\newcommand{\dd}{\partial}
\newcommand{\bftau}{\mbox{\boldmath$\tau$}}
\newcommand{\bfo}{\mbox{\boldmath$\omega$}}

\newcommand{\half}{\frac{1}{2}}

\def\dal{\,\lower0.3ex\vbox{\hrule\hbox{\vrule\kern2pt\vbox{\kern4pt\kern4pt}
\kern2pt\vrule}\hrule}\,}

\def\s{\sigma}
\def\o{\omega}

\begin{document}

\title{{\bf THE SKYRMION IN THE NUCLEUS}}
\vspace{1 true cm}
\author{G. K\"ALBERMANN \\
Faculty of Agriculture \\
and \\
Racah Institute of Physics \\
Hebrew University, 91904 Jerusalem, Israel \\}

\maketitle

\begin{abstract}
\baselineskip 1.5 pc
The energy levels of a skyrmion in nucleus are calculated 
in a field theory of skyrmions 
coupled to the dilaton field and the $\o $ meson
. The central potential fits well with expectations. The nucleon spin-orbit 
interaction derived from the omega meson in a rotating frame gives the correct
level splittings. The same interaction originating from the Thomas precession
effect is negligible. Energy levels are calculated for closed
shell nuclei. The meson fields are obtained from a Thomas-Fermi mean field
approximation to the nucleus.
\end{abstract}

\newpage
\baselineskip 1.5 pc

One of the key successes of conventional nuclear theory was the
ability to describe in a simple manner the shell structure of nuclei.
Harmonic oscillator, or Woods-Saxon potentials give the gross features
of the single particle levels of nucleons. In the present work we address
the question of the suitability of the Skyrme model in achieving the same goal.

In previous publications\ci{kal1,kal2} a model of a fluid of skyrmions
in nuclear matter and finite nuclei was developed in a dilute fluid 
approximation.
The skyrmions were regarded as essentially free objects interacting via the
dilaton field and the $\o$ meson in a mean field approach.
The model succeeds in reproducing the main features of nuclear matter as well as
closed shell nuclei. It incorporates scale and chiral invariance. 
The former is manifestly broken by a potential inspired in the trace anomaly.
In the present work we take advantadge of the scalar fields derived
from the mean field approximation in order to generate the central and 
spin-orbit
interactions of a single skyrmion in the nucleus.

In ref.~\ci{kal1,kal2} a considerable effort was invested in generating
the potential for the dilaton in order to reproduce the main features
of nuclear matter and finite nuclei.
As it is our aim now
to focus on the single-particle energy leveles of the nucleon in the nucleus,
we will simplify the calculation by freeing ourselves from a specific 
parametrization for the dilaton potential. Instead, the measured nuclear
densities will be used to produce the mean fields in a Thomas-Fermi 
approximation.
A more consistent approach would be to use a Hartree (or Hartree-Fock)
method. However, the goal of the present work is to see whether we can reproduce
the key elements of the nucleon binding energy, especially the spin-orbit
interaction and not the fine details. 
In the Walecka-type models\ci{ser1,ser2}, this potential arises
from relativistic effects in the Dirac equation. While the central potential
is due to the subtraction of scalar and vector interactions, the
spin-orbit force originates from their addition. The scalar and vector
potentials being large, produce a sizeable spin-orbit force.

We here investigate various possible sources of spin-orbit interactions and
find that there is a large potential due to the motion of the skyrmion in
a quiescent nucleus. Viewed from the frame of the nucleon the nucleus flows
past it. The $\o$ meson in this frame develops a spatial component that
is responsible for the spin-orbit coupling. This is a specific feature of
the skyrmion picture, as evidenced by the presence of the skyrmion moment of
inertia in the interaction. This would be senseless in the Dirac approach for
which the nucleon is pointlike.

The lagrangian adopted for the model is

\bea \label{skydil}
 {\cal L} & = &  \e^{2\s}\bigg[\half~\Gamma_0^2\,\dd_\mu\s\,\dd^\mu\s
- {F_\pi^2 \over 16}\tr(L_\mu L^\mu)\bigg]
+{1 \over 32 e^2}\tr[L_\mu,\, L_\nu]^2 \nono
&-& V_{\s}- {1\over 4} {G_{\mu\nu}G^{\mu\nu}}
+ \half~e^{2\s} m_\o^2~~\o_\mu^2 -g_V~~ \o_{\mu} B^{\mu}
\eea
Here
\be \label{Lmu}
L_\mu \equiv U^\da\dd_\mu U,
\ee
where $U({\bf r},t)$ is the chiral field, $F_{\pi}$ is the pion decay constant 
$e$ the Skyrme parameter, and
\bea \label{rho}
G_{\mu\nu} = \dd_{\mu}  \o_{\nu} - \dd_{\nu}  \o_{\mu}
\eea
Following ref.~\ci{kal1} the ans\"atze for the scalar
and vector fields become

\bea \label{ansatz}
U(\bf r) & = & \exp[i\bftau\cdot\hat{\bf r} F(r)], \nono
\o^{\mu}(\bf R) & = & \delta_{{\mu}0}~\o(R)\nono
\eea
where ${\bf R}$ measures the distance from the center of the nucleus at rest,
and ${\bf r}$ is the coordinate from the center of the skyrmion.

Averaging over the coordinates $R$ and the corresponding momenta with 
the distribution function for zero temperature, the energy of the 
nucleus with spherical symmetry reads
\bea \label{energy}
E & = & 4 {\pi} \int{ R^2 dR~~E(R)}\nono
E(R) & = & E_{\s} + E_{\o} + E_{int} + E_{sk}
\eea
where
\bea \label{detail}
E_{\s} & = & \half~\Gamma_0^2~e^{2\s}\s'^2 + V_{\s}  \nono
E_{\o} & = & -\half~\o'^2-e^{2\s}\frac{m_\o^2\o^2}{2}\nono
E_{int} & = &  g_V~\o~B \nono
E_{sk} & = & \frac {2}{\pi^2} \int_0^{k_F} k^2 dk \sqrt{k^2 + M^2}
\eea
where $k_F$ is the nucleon local, $R$ dependent, 
Fermi momentum and,
\bea \label {mass}
 M(R) & = & 4\pi\int_0^\infty r^2\,dr M(r)\nono
M(r) &= &\e^{2 \s}\frac{F_\pi^2}{8}
\left[F'^2 + 2 \frac{\ssF}{r^2}\right] + \frac{1}{2 e^2}
\frac{\ssF}{r^2} \bigg[\frac{\ssF}{r^2}+2
F'^2\bigg]
\eea
is the nucleon mass. Using the virial theorem, or the skyrmion equations of
motion, the mass scales as \ci{kal1}
\be \label{mascale}
M(R) = e^{\s} M_0
\ee
where $ M_0$ is the skyrmion mass for $\s$ = 0.
In eq.~(\ref{detail}) primes denote derivatives with 
respect to $R$, whereas in eq.~(\ref{mass}) they represent derivatives with 
respect to $r$.
Also
\bea \label {density}
B(R) = \frac {2~k_F^3}{3~{\pi}^2} 
\eea
The Euler-Lagrange equations for the mean fields become

\bea \label {equations}
{\Gamma}_0^2 ~~ e^{2\s} \bigg(\s'' + 2\s'^2  +  \frac{2\s'}{R} \bigg)  -
\frac{dV_{\s}}{d\s} +m_{\o}^2~\o^2~ e^{2\s} - \frac{\dd E_{sk}}{\dd\s}= 0 \nono
\o'' + \frac {2 \o'}{R}  - m_{\o}^2~\o~e^{2 \s} + g_V~B = 0 \nono
\eea
The ground state of the nucleus for fixed number of nucleons
is obtained by minimizing the energy, constrained by means of a Lagrange
multiplier, with respect to the local wavenumbers $k_F$ (no distinction
is made here between protons and neutrons)

The algebraic equations for the multipliers become \ci{ser1}

\bea \label {mu}
\mu & = & g_V~\o + \sqrt{k_F^2 + M^2}
\eea
We solve the equation for the $\o$ meson of eq.~(\ref{equations}) with $B$ 
replaced by the measured densities and the dilaton determined by the
chemical potential. In this manner we avoid tedious parametrizations of
the dilaton potential in order to fit the densities. The 
assumption is that such a potential exists, and the results of ref.~\ci{kal1}
support it.

A skyrmion in the nucleus, will be affected by the $\s$ and $\o$ mean
fields. These will produce the central potential. In order to uncover
the spin-orbit interaction we consider three types of contributions.
A Lorentz boost of the skyrmion, the Thomas precession induced by
two succesive Lorentz transformations and, dynamical $\o$ meson
effects.
The first contribution may be found by boosting rigidly the whole skyrmion
by means of the transformation of the argument of the 'hedgehog' of 
eq.(~\ref{ansatz}) with velocity $\bf v$
\bea \label {boost}
{\bf{r}}~\rightarrow~{\bf{\tilde r}} = {\bf r} + {\bf r}\cdot{\bf{\hat v}}
~{\bf{\hat v}}~(\gamma-1) -\gamma~\bf R
\eea
where $\gamma = \frac{1}{\sqrt{1-v^2}}$, together with a rigid rotation of
the skyrmion with collective coordinate matrices $A$\ci{witten}
\bea \label {collective}
U({\bf{\tilde r}})~\rightarrow~ A(t) U({\bf{\tilde r}}) A^{\da}(t)
\eea
Inserting the transformations in the skyrmion lagrangian of eq.~(\ref{skydil})
and after a lengthy calculation it is found that the spin-orbit potential
from this transformation vanishes.

A second possible contribution to the spin-orbit interaction comes from the
well-known Thomas precession that arises from two consecutive Lorentz 
transformations.
A way to implement this transformation consists in considering the isospin
vector matrix as time dependent

\bea \label{Thomas}
 {\dot{\bftau}} = -{\bf \Omega}_T~\bf x~\bftau
\eea
where ${\bf \Omega}_T$ is the Thomas frequency.
Again inserting in the skyrmion lagrangian of eq.~(\ref{skydil}) the spin-orbit
interaction to lowest order in the velocity is found to be 
\be\label{so1}
U_{s.o.} \approx-\frac{{\bf S}\cdot{\bf L}}{2~M_0^2~R} \frac{\dd V_C}{\dd R}
\ee
where $V_C$ is the central potential of the skyrmion in the nucleus, $\bf S$
is the spin and $\bf L$ the angular momentum. The same result as in
standard textbook derivations\ci{jackson}. 
In eq.(\ref{so1}) we have used the projection formula \ci{witten}

\bea\label{proj}
{\dot A^{\da}}~A = \frac{-i~\bftau\cdot\bf S}{2~\lambda(R)}
\eea
where $\lambda(R)$ is the moment of inertia of the nucleon\ci{witten}
\bea\label{lambda}
\lambda(R) & = & \frac{2 \pi}{3}\int~r^2~dr~\Lambda(R)\nono
\Lambda(R) & = & sin^2(F)\bigg[F_{\pi}^2~e^{2\s}~+~\frac{4}{e^2}
\bigg(F'^2+\frac{sin^2(F)}{r^2}\bigg)\bigg]
\eea
where F is the skyrmion profile of eq.~(\ref{ansatz}) and the $R$ dependence
enters through the dilaton field $\s$.
It will turn out, as expected beforehand, that the spin-orbit of eq.~(\ref{so1})
is quite negligible, due to the $\frac{1}{M_0^2}$ dependence.

The third and most important possible source of spin-orbit force is due to the
transformation of the static scalar fields to a rotating frame.
In order to find this potential, it is first desired to find the mean fields
for a streaming nucleus. In this case there arises a spatial component of the
$\o$ meson field. An appropriate approximate 
ansatz for this  component is \ci{ser1}
\bea\label{omega}
{\bfo} = {\bf V}~\o_1(R) = ({\bf \Omega~x~R})~\o_1(R)
\eea
where $\bf V$ is the tangential velocity of the nucleus at each $R$ and 
$\bf\Omega$
the angular velocity. At the same time the nucleus baryon density develops
a time dependent piece (in a nonrelativistic approximation) 
of the form \ci{riska}
\bea\label{b}
{\bf B}({\bf R}) = {\bf V} B_0(R)
\eea
where $B_0(R)$ is the static baryon density of the nucleus.  
Inserting eqs.(\ref{omega},\ref{b}) above  in the lagrangain 
(\ref{skydil})
with a distribution function for a nucleus at zero temperature
and averaging over the angular directions \ci{ser1, kal1} 
of $\bf R$ we find the equation of motion for $\o_1$ to be
\bea\label{omega1}
\o_1'' + \frac {4 \o_1'}{R}  - m_{\o}^2~\o_1~e^{2 \s} + g_V~B = 0 \nono
\eea
This is very similar to the equation of motion of the static $\o$ in 
eq.~(\ref{equations}).
We solve equation (\ref{omega1}) for each nucleus using the
dilaton field of the static case and demanding a vanishing
$\o_1$ at infinity.

In order to find the corresponding spin-orbit interaction we consider
a nucleon spinning at rest with a nucleus rotating with an velocity
${-\bf V}(R)$ oposite to the direction of rotation of the nucleon. Using
the collective coordinate quantization scheme of eq.~(\ref{collective}),
and the projection formula of eq.~(\ref{proj}) we find
\bea\label{uso2}
W_{s.o.} = \frac{-\bf S \cdot \bf L}{2~M_0~\lambda(R)}~\o_1(R)
\eea

It is clear that $W_{s.o.}$ is more important than 
$U_{s.o.}$ due to the $\frac{1}{M_0}$ dependence.
It is also a purely skyrmion spin-orbit as evidenced by the presence
of the moment of inertia in the potential. In the usual Dirac type of
Walecka models \ci{ser1}, the spin-orbit interaction arises from
the coupling to the lower components of the Dirac wave function, whereas
here it arises from the interaction of the rigid rotation of the nucleon
with the flow of the mean fields.

The Hamiltonian of a single Syrmion embedded in a nucleus becomes
\be\label {ham}
H = \sqrt{p^2 + M^2}+ g_V~\o + W_{s.o.} + U_{s.o.} 
\ee
where $p$ is the nucleon momentum, the conjugated variable to the skyrmion
center location $R$.

Quantizing the coordinate $R$ we obtain an effective Schr\"odinger equation
for the radial wave function of the skyrmion center with total energy $E$
 
\bea\label{schroe}
\bigg[\frac{\dd^2}{{\dd R}^2}+\frac{2~\dd}{\dd R}
-\frac{l~(l+1)}{R^2}-Q(R)\bigg]{\Psi}= 0\nono
\eea
where
\bea\label{q}
Q(R)  =  e^{2\s}~M_0^2-{\bigg(-g_V~\o +~E~ -W_{s.o.}\bigg )}^2-Z(R)\nono
Z(R)~\approx\frac{g_V}{2 M(R)}\bigg[\frac{{\dd}^2}{{{\dd}R}^2}
+2\frac{\dd}{\dd R}\bigg]\o+ 2~M_0~U_{s.o.}
\eea
where we have expanded the square root in eq.~(\ref{ham}) when operating on the
potential to order $p^2$. The central potential entering the calculation of
$U_{s.o.}$ of eq.(~\ref{so1}) is given by
\bea\label{vc}
V_C = \frac{Q_1(R)}{2 M_0}
\eea
where $Q_1$ is given by $Q(R)$ of eq.(\ref{q}), but without the spin-orbit 
pieces.

We have solved the Schr\"odinger equation (\ref{schroe}) for the ground state
single particle levels of the magic nuclei
$C^{12} , O^{16}$ and $Ca^{40}$.
Table 1 shows the comparison between the predicted binding energies
the experimental ones \ci{fuchs,landaud},
averaged over proton and neutron states. 
The results show that the skyrmion picture of both the central 
and the spin-orbit interaction is quite good.
The spin-orbit originates solely from the $\o$ meson as viewed by the
rotationg skyrmion, in contradistinction to the Dirac case in which
both scalar and vector fields act together to produce a large 
interaction.
\vspace{3 pc}

\begin{table}
\caption {{\bf binding energies of single particle levels}}
	\begin{center}
         \medskip
         \begin{tabular}{|c|c|c|c|}
           \hline
Nucleus&Shell&calculated energy&experimental energy\\
&&MeV&MeV\\
&&&\\
           \hline
&&&\\
$C^{12}$&$1s_{\frac{1}{2}}$&36.3&35.2\\
&1$p_{\frac{3}{2}}$&15.7&16.9\\
&&&\\
           \hline
&&&\\
$O^{16}$&$1s_{\frac{1}{2}}$&37.1&43$\pm$5\\
&$1p_{\frac{3}{2}}$&20.5&20.1\\
&$1p_{\frac{1}{2}}$&15.6&13.9\\
&&&\\
           \hline
&&&\\
$Ca^{40}$&$1s_{\frac{1}{2}}$&48&50$\pm$10\\
&$1p_{\frac{3}{2}}$&35&34$\pm$6\\
&$1p_{\frac{1}{2}}$&30.7&34$\pm$6\\
&$1d_{\frac{5}{2}}$&21&18.5\\
&$2s_{\frac{1}{2}}$&15.7&14.5\\
&$1d_{\frac{3}{2}}$&14.8&12\\
&&&\\
           \hline
         \end{tabular}
     \end{center}
\end{table}

\end{document}